\documentclass[%
 reprint, 
 amsmath,amssymb,
 aps, physrev,
]{revtex4-2}

\usepackage{graphicx}
\usepackage{dcolumn}
\usepackage{bm}

\begin{document}


\title{\textbf{Experimental demonstration for precisely tuning the focal length of finite-aperture focused beams and vortex.} 
}%

\author{Shiyu Li}
\affiliation{State Key Laboratory of Ocean Engineering, School of Ocean and Civil Engineering, Shanghai Jiao Tong University, Shanghai, 200240, China}

\author{Yicheng Feng}
\affiliation{State Key Laboratory of Ocean Engineering, School of Ocean and Civil Engineering, Shanghai Jiao Tong University, Shanghai, 200240, China}


\author{Weiwei Cui}
\affiliation{State Key Laboratory of Precision Measuring Technology and Instruments, Tianjin University, Tianjin 300072, China}

\author{Zhixiong Gong}
\email{Corresponding author: zhixiong.gong@sjtu.edu.cn}
\affiliation{State Key Laboratory of Ocean Engineering, School of Ocean and Civil Engineering, Shanghai Jiao Tong University, Shanghai, 200240, China}
\affiliation{Key Laboratory of Marine Intelligent Equipment and System, Ministry of Education, Shanghai, 200240, China}


\date{\today}

\begin{abstract}
High-frequency focused ultrasound is widely used in biomedical applications such as high-resolution imaging, neuromodulation, particle manipulation, and so on. 
However, dynamic tuning of the focal plane in conventional systems often relies on mechanically adjustable components or array-based control with complex system and high cost. 
In this work, an optically transparent, planar compact piezoelectric ultrasonic transducer was designed and fabricated by truncating an ideal spherical wavefront with a plane, enabling high-frequency focused ultrasound generation and convenient integration with microscopic platforms. The acoustic field was characterized experimentally at the focal plane under the design frequency and at propagation planes near the design frequency to evaluate the focal tuning. An approximate linear relation between the focal length and driving frequency near the design one is derived theoretically, and the finite-range tuning behavior is interpreted using the stationary-phase condition. Both theory and experiment show that the focal length varies approximately linearly with excitation frequency near the design frequency. Water-tank measurements agree well with the theoretical prediction, confirming the proposed model. This work provides a simple and cost-effective approach for focal tuning in compact high-frequency ultrasound devices. 
\end{abstract}

\maketitle


\section{\label{sec:Introduction}Introduction}

Focused ultrasound fields play a central role in a wide range of applications, including high-resolution imaging~\cite{foster2000advances}, noninvasive therapy~\cite{ter2001high,izadifar2020introduction} and particle manipulation~\cite{li2026review,baudoin2020acoustic,baudoin2020spatially}. Conventional focused beams generated by curved transducers~\cite{malietzis2013high,zhang2024design} or phased arrays~\cite{azar2000beam,hynynen2016image}, enable strong energy localization and have been extensively studied. Beyond these axisymmetric fields, structured acoustic waves such as acoustic vortices offer an additional degree of freedom through their orbital angular momentum, enabling advanced functionalities such as selective trapping and rotational manipulation~\cite{li2026review}. For both conventional focused beams and acoustic vortices, precise control of the focal position is essential for achieving reliable acoustic focusing tailored to specific biomedical applications~\cite{kim2022focal,zhang2025knn}.

In recent years, planar and compact focused acoustic transducers have emerged as an attractive alternative to conventional transducer~\cite{baudoin2019folding}. This approach is particularly compatible with microfabrication techniques and enables the realization of interdigitated electrode patterns for generating both focused beams~\cite{gong2022single,li2025reversing} and focused vortex~\cite{baudoin2020spatially,gong2021three} on planar substrates. 
Dynamic control of the focal position is commonly implemented either by mechanically tunable acoustic elements, such as variable-focus acoustic lenses~\cite{song2013acoustic,donahue2014experimental}, or by electronically controlled phased arrays~\cite{kujawska2017annular} that steer and refocus the beam through prescribed phase delays. Although effective, these approaches generally require additional structural or driving complexity and may increase system cost and power consumption, which can be disadvantageous for compact or integrated implementations~\cite{el2011driving}. An alternative strategy is to exploit the intrinsic frequency dependence of the focusing condition. By varying the driving frequency in the vicinity of the design frequency, the focal length can be dynamically adjusted, thereby opening up new possibilities for applications such as the axial manipulation of microparticles~\cite{baudoin2019folding,gong2021three}.

Frequency-dependent focal tuning has been reported in several classes of acoustic devices. For focused vortex devices, tunable focusing has been demonstrated in active spiral Fresnel zone plates~\cite{muelas2020active} and further analyzed theoretically in frequency-tuned spiraling acoustical tweezers~\cite{gong2021three}. In contrast, for conventional focused beams, frequency-controlled axial focal shift has also recently been discussed in focused beam transducers~\cite{li2025reversing}. However, two important gaps remain. First, concentric-ring focused beam devices and Archimedes-Fermat spiral focused vortex devices have so far been studied largely as separate cases, and a unified theory-simulation-experiment comparison of their frequency-dependent focal tuning characteristics is still lacking. Second, although focal shift under frequency detuning has been observed and modeled in specific devices, a compact analytical expression describing the approximately linear variation of focal length in the vicinity of the design frequency has not yet been established in a common framework. Moreover, the physical origin of the finite frequency tuning range over which effective focusing can be maintained remains to be clearly elucidated.

In this work, we present a unified theoretical, numerical, and experimental study of frequency-dependent focusing in planar finite aperture acoustic transducers for both conventional focused beams and first-order acoustic vortex. The transducer designs are obtained by truncating the wavefront of an ideal spherical wave (for the focused beam) and that of a spherical vortex (for the focused vortex), resulting in concentric circular and Archimedes-Fermat spiral electrode, respectively. On this basis, we derive an analytical relation between focal length and excitation frequency, and further explain the underlying mechanism of frequency tuning focal length variation. The simulated predictions are validated by angular spectrum method and water tank measurements using a scanning hydrophone. The measured acoustic fields and extracted focal positions agree well with the simulations, confirming the predicted focal length and its near-linear dependence on frequency around the design frequency. This work provides a simple and predictive route for focal tuning based solely on frequency control and may be useful for compact acoustic devices in applications such as acoustic tweezers, microfluidics, and biomedical ultrasound.

The paper is orginized as follows. Section \ref{sec: theory} describes the characteristics of focused beams and focused vortex, as well as their associated electrode trajectory functions. It further derives the theoretical expression describing the dependence of the focal length on frequency.
Section \ref{sec: Numerical method} outlines the angular spectrum method and the corresponding numerical scheme.
In Section \ref{sec:experiment}, we describe the MEMS (i.e., Micro-Electro-Mechanical System) fabrication process for focused ultrasound transducers and the water tank scanning method for acoustic field measurement. Section \ref{sec:results and discussion} compares the simulated and measured results for focused vortex and focused beam, validates the frequency tuning focal adjustment capability of the fabricated device.  Furthermore, the observed behavior of frequency-tuning focal length variation is explained, with excellent agreement demonstrated among theory, simulation, and experiment.
A summary of key findings is provided in Section~\ref{sec:conclusion}. 


\begin{figure*}
\includegraphics[width=14.6cm]{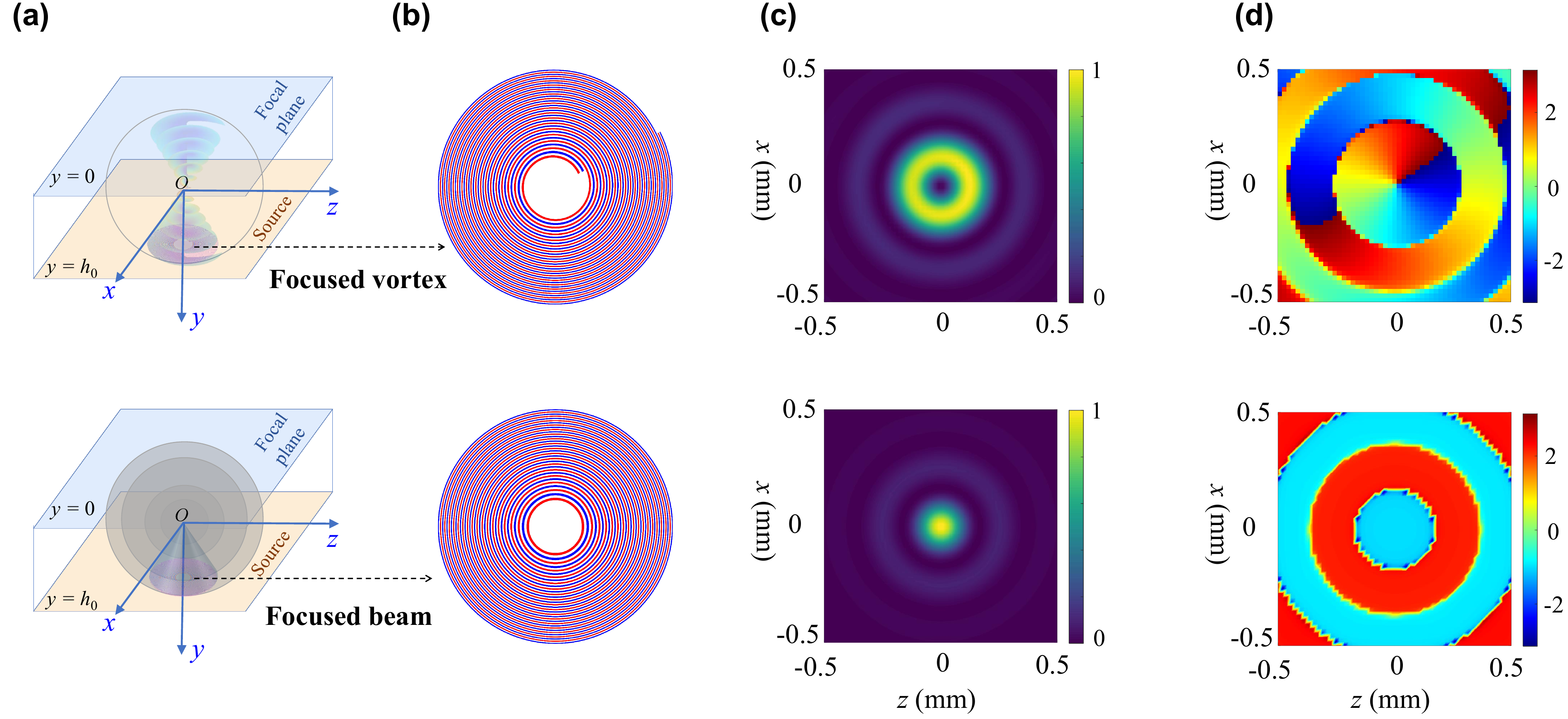}
\caption{(color online) Comparison of the focused vortex and focused beam generated by planar interdigital transducers. (a) Plane truncation of ideal spherical vortex (top) and spherical wave (bottom) with sources placed all over the $4\pi$ steradians to produce finite-aperture devices for cleanroom fabrication and practical applications. 
(b) Electrode patterns of focused vortex (top) and focused beam (bottom). There is a phase shift of $\pi$ between the blue and red (excited as positive and negative in experiments) electrodes. 
(c) Theoretical results of the normalized acoustic pressure magnitude in the focal plane for the focused vortex (top) and focused beam (bottom). 
(d) The corresponding phase distributions in the focal plane. }
\label{Fig1}
\end{figure*}

\section{\label{sec: theory} Theoretical model}
\subsection{\label{sec: Beam model}  Finite Aperture Focused Beam Model}
In this work, both the focused beam and focused vortex transducers were designed with the same parameters, namely a center frequency of 5 MHz ($f_0$ = 5 MHz), a focal length of 6 mm ($h_0$ = 6 mm), and 17 electrode turns ($N = 17$).
The design of the planar transducers considered in this work is based on the phase distribution of a converging acoustic field (see Fig.\ref{Fig1}). The ideal converging Hankel spherical vortex representation can be written as~\cite{baudoin2019folding,baudoin2020spatially,gong2021three}
\begin{equation}
p_1^*(r, \theta, \varphi,t)=p_0 h_n^{(2)}(k r) P_n^m(\cos \theta) e^{i(m \varphi-\omega t)},
\end{equation}
where $p_0$ is the pressure amplitude, $h_n^{(2)}$ is the spherical Hankel function of the second kind, $P_n^m(\cos \theta)$ is the associated Legendre function, $k$ is the acoustic wavenumber, and $m$ denotes the topological charge. The azimuthal phase factor $e^{i m \varphi}$ introduces a helical phase distribution and gives rise to a phase singularity on the beam axis, which is the defining feature of the acoustic vortex~\cite{baresch2013spherical,baudoin2020acoustic}.

To generate such a field using the planar transducer, the electrode pattern must reproduce the phase distribution of the converging spherical vortex wave truncated by a plane as shown in Fig.\ref{Fig1}(a)~\cite{baudoin2019folding}. The required phase on the transducer surface can therefore be obtained from the phase of the spherical vortex field.
By imposing a phase difference of $\pi$ between adjacent electrodes of the interdigital transducer, the electrode trajectories corresponding to the vortex phase distribution can be derived. The resulting electrode radii can be written as ($m$=1)~\cite{baudoin2019folding,gong2021three}
\begin{equation}
\begin{aligned}
& R_1=\frac{1}{k} \sqrt{\left(\varphi+C_1\right)^2-(k h_0)^2} \\
& R_2=\frac{1}{k} \sqrt{\left(\varphi+C_1+\pi\right)^2-(k h_0)^2}
\end{aligned}
\end{equation}
where $C_1$ is a constant and $h_0$ is the focal length.
These equations describe Archimedes-Fermat spiral electrode trajectories (see Fig.\ref{Fig1}(b), top) that impose the required helical phase distribution on the emitted acoustic wave, thereby producing a focused acoustic vortex (see Fig.\ref{Fig1}(c) and Fig.\ref{Fig1}(d), top).

The conventional focused beam corresponds to the non-vortex, axisymmetric limit of the above formulation. When the azimuthal phase dependence vanishes ($m$=0), the helical phase term disappears and the acoustic field becomes axisymmetric. In this case, the vortex formulation reduces to a non-vortex converging field without angular phase variation.
In the present work, this non-vortex configuration (ideal converging spherical focused beam) is described by the spherical wave model in the spherical coordinates ($r$,$\theta$,$\varphi$)~\cite{gong2022single,li2025reversing}

\begin{equation}
p_0^*(r, \theta, \varphi, t)=p_0 e^{i(k r-\omega t)} / r,
\end{equation}
 This expression represents a converging spherical wave that forms a focused beam. The radial positions of the electrodes can be written as~\cite{gong2022single,li2025reversing}
\begin{equation}
\begin{aligned}
& R_1=\frac{1}{k} \sqrt{(C_0+2 n \pi)^2-(k h_0)^2} \\
& R_2=\frac{1}{k} \sqrt{[C_0+(2 n+1) \pi]^2-(k h_0)^2}
\end{aligned}
\end{equation}
where $C_0$ is a constant, $n$ is an integer, and each value of $n$ corresponds to a set of two electrodes.
These curves define the trajectories of the interdigital electrodes that generate the focused acoustic beam (see Fig.\ref{Fig1}(b)-(d), bottom).
It should be noted that in all electrode configurations, the electrode width is chosen to be half of the spacing between two adjacent electrodes of opposite polarity.

\subsection{\label{sec: Analytical equation}  Analytical Expression For Focal Length Versus Frequency} 
Based on the theoretical derivation method for focused beams, we extend it to the case of focused vortex~\cite{li2025reversing}. The relation between focal length and excitation frequency can be derived directly from the transducer aperture $R_{max}$. For both the focused beam and the focused vortex, the outermost active electrode branch defines the transducer aperture, which can be written in the unified form
\begin{equation}
    R_{max}=\frac{1}{k} \sqrt{2 k h_0 C_{m,max} +C_{m,max}^2}
\end{equation}
where $C_{m,max}$ is a geometry-dependent constant, $h_0$ is the focal length, and $k=2 \pi f/c$ is the wavenumber. 

Since the transducer geometry is fixed after fabrication, $R_{max}$ remains unchanged when the excitation frequency varies. Therefore, equating $R_{max}$ at the design frequency $f_0$ and the arbitrary frequency $f$ yields
\begin{equation}
h_m(f)=\alpha h_0+\frac{C_{m,max}}{2 k_0}\left(\alpha-\frac{1}{\alpha}\right),  \alpha=f/f_0
\label{Eq:Focal length versus frequency}
\end{equation}
Here, $h_0$ and $k_0$ denote the focal length and wavenumber at the design frequency, respectively. Near the design frequency, this relation may be linearized, showing that the focal length varies approximately linearly with excitation frequency $f$.
When the excitation frequency $f$ is close to the design frequency $f_0$, i.e.,$\alpha \approx 1 $, Eq.\eqref{Eq:Focal length versus frequency} can be linearized as

\begin{equation}
h_m(f) \approx (h_0 + \frac{C_{m,max}}{k_0} )\alpha-\frac{C_{m,max}}{k_0}
\label{Eq:linear equation}
\end{equation}
For the two transducers investigated in this work, namely the focused beam and the first-order focused vortex, the number of electrode turns is the same ($N=17$). As a result, the geometry-dependent coefficient takes the identical value $C_{0,max}=C_{1,max}=(2N+1)\pi=35 \pi$.
Based on the inverse function of Eq.\eqref{Eq:Focal length versus frequency}, the required driving frequency can be determined from a specified target focal length and number of transducer turns.


\section{\label{sec: Numerical method} Numerical Simulation}
\subsection{Angular Spectrum Method}
Given that this study focuses on the spatial acoustic field generated by a finite aperture focused transducer, the angular spectrum method (ASM) was employed for comparative simulations.  The angular spectrum method decomposes a complex acoustic field into a series of plane waves propagating in different directions, thereby enabling the calculation of the acoustic field at an arbitrary plane~\cite{gong2021three,gong2022single,li2025reversing}.  In this section, we briefly review the application of the angular spectrum method to the calculation of acoustic propagation. To maintain consistency with the coordinate system of the experimental setup (see Fig.~\ref{Fig3}), the beam axis is taken to be along the $y$-direction.  Assuming the incident sound pressure field is known at the source plane, $p_{in}|_{y=0}=p_{in}(x,0,z)$, and the angular spectrum $S(k_x,k_z)$ of the source plane is then obtained via a 2D Fourier transform
\begin{equation}
\begin{aligned}
S\left(k_{x}, k_{z}\right)=\int_{-\infty}^{+\infty} \int_{-\infty}^{+\infty}  p_{in}(x, 0, z) e^{-i k_{x} x - i k_{z} z} d x d z
\end{aligned},
\label{S(kx,ky)}
\end{equation}
where $ k_{x} $ and $ k_{z} $ are the wavenumber vector components of the wave vector $\boldsymbol{k}$.
Since each angular spectral component $S(k_x,k_z)$ represents a plane wave, the propagation of the sound field between different planes can be calculated using plane wave properties as follows:
\begin{equation}
\left.S\left(k_x, k_z\right)\right|_{y=y_s}=\left.S\left(k_x, k_z\right)\right|_{y=0} e^{i k_y y_s}
\end{equation}
where $y_s$ is the axial distance from the source plane ($y=0$) to the target plane ($y=y_s$), and $k_y=\sqrt{k^2-k_x^2-k_z^2}$ is the $y$-component of the wave vector. Following the steps above, the sound field at any target position can be calculated as:
\begin{equation}
\begin{aligned}
p_{i n}(x, y, z)= & \left.\frac{1}{4 \pi^2} \iint_{k_x^2+k_z^2 \leq k^2} S\left(k_x, k_z\right)\right|_{y=0} \\
& \times e^{i k_x x+i k_z z+i \sqrt{k^2-k_x^2-k_z^2} y} d k_x d k_z .
\end{aligned}
\label{Eq:ASM}
\end{equation}
Here, the integration is confined to the region $k_x^2+k_z^2 \leq k^2$, so as to eliminate the influence of evanescent waves.
By setting $y=h_0=6$~mm (the focal length) in Eq.\eqref{Eq:ASM}, the acoustic field distribution at the focal plane can be obtained.  

\subsection{Numerical Implementation}
The numerical calculations were implemented in MATLAB using a self-developed code based on the angular spectrum formulation. The acoustic pressure distribution on the transducer surface was defined according to the designed phase pattern of the interdigital electrodes. The acoustic field was then propagated numerically along the axial direction to obtain the three-dimensional pressure distribution.

To investigate the frequency-tunable focusing behavior, simulations were performed for a set of excitation frequencies around the design frequency. For each frequency, the acoustic pressure field was calculated in a region covering the focal area.

The focal length was extracted from the simulated acoustic fields using criteria adapted to the topology of the generated field. For the focused beam, the focal position was determined from the axial pressure amplitude distribution, and the focal length was defined as the axial location where the pressure amplitude on the beam axis reached its maximum. For the focused vortex, however, the on-axis pressure vanishes in the vicinity of the beam center because of the phase singularity, so the focal plane cannot be identified from the axial response alone. Instead, a main-ring energy criterion was adopted. At each axial position, the radius of the dominant annular lobe was first determined from the radially averaged intensity profile, and the acoustic energy contained within a narrow annular band centered at this radius was then evaluated. The focal length was finally defined as the axial location at which this ring-band energy attained its maximum. In this way, the focal position extraction remains consistent with the distinct field characteristics of the focused beam and the focused vortex.

\section{\label{sec:experiment} Fabrication and experimental methods}
\subsection{MEMS Transducer Fabrication}
\begin{figure}
\includegraphics[width=0.45\textwidth]{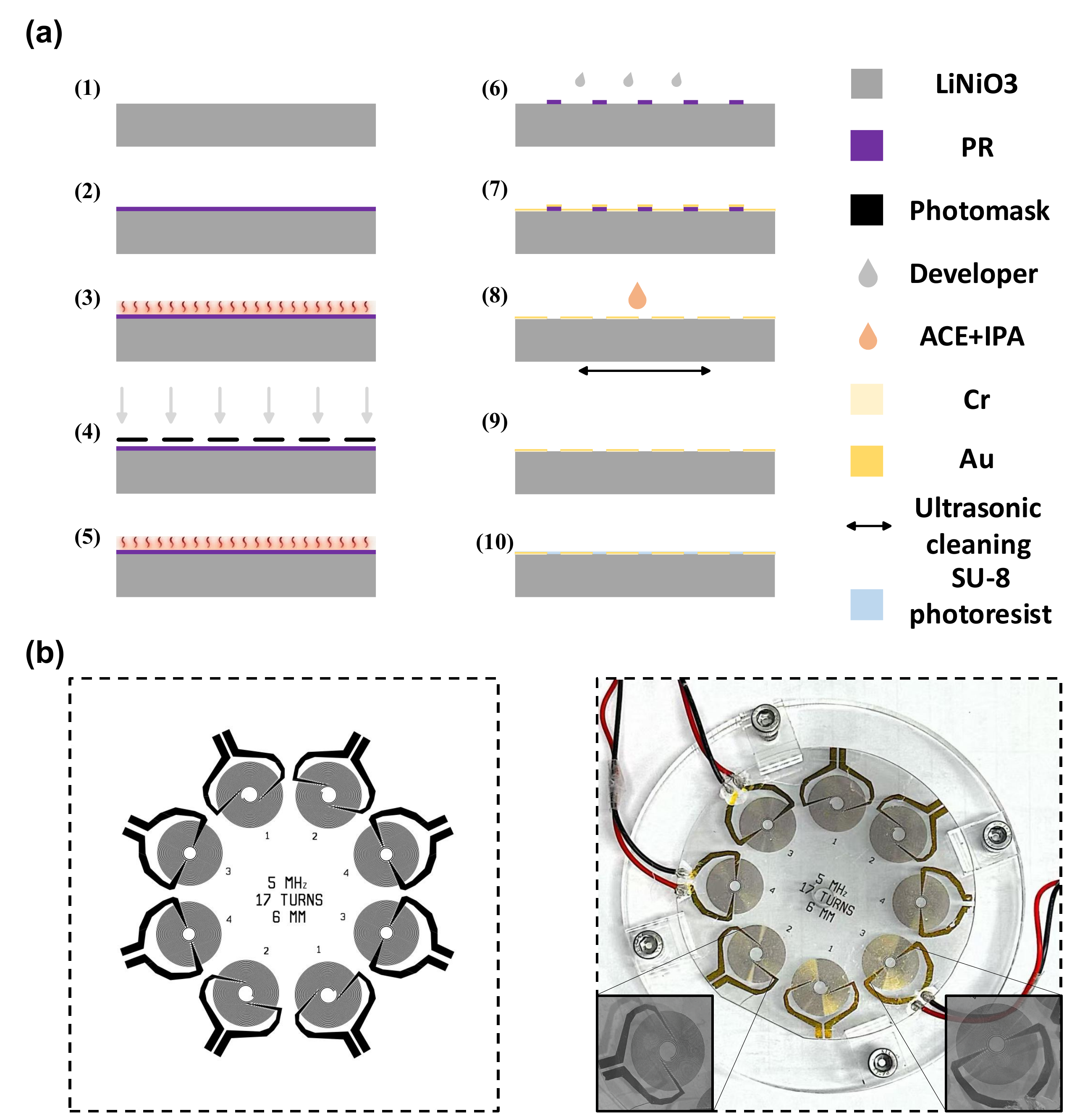}
\caption{\label{Fig2:MEMS} (color online) Schematic of MEMS transducer fabrication. (a) Schematic diagram of the MEMS fabrication procedure. 
 (1) Cleaning. (2) Spin coating. (3) Soft baking. (4) Exposure. (5) Post-exposure baking. (6) Development. (7) Magnetron sputtering. (8-9) Lift-off. (10) Water sealing. Note: Colors and scale in the figure are schematic representations only, not representing actual conditions. (b) Mask design (left) and device photograph after waterproof packaging and wire bonding (right). The two enlarged subfigures show the actual photographs of the focused vortex and the focused beam, and the bottom shadow is due to photography. Note: Devices 1 and 2 are focused vortex devices, devices 3 and 4 are focused beam devices. }
\end{figure}

The planar focused ultrasonic transducer was fabricated using standard MEMS processes (see Fig.\ref{Fig2:MEMS}(a)). The piezoelectric substrate is 0.5-mm-thick 36$^\circ$ Y-cut lithium niobate (LiNbO3), whose optical transparency enables convenient microscopic inspection. The process begins with substrate cleaning using acetone and isopropyl alcohol in an ultrasonic bath, followed by drying with nitrogen ($N_2$). Subsequently, HMDS treatment is applied to promote adhesion, and AZ5200 photoresist is spin-coated onto the substrate to a thickness of approximately 3 $\mu$m. After soft baking, UV lithography was performed using an optical mask (see Fig.\ref{Fig2:MEMS}(b)) to transfer the electrode pattern onto the photoresist. Following post-exposure baking and development, oxygen plasma treatment was applied, and then a Cr/Au (40nm/400nm) film was deposited on the lithium niobate substrate by magnetron sputtering. Finally, a lift-off process was performed by ultrasonic cleaning in acetone and isopropyl alcohol to remove the photoresist and unwanted metal layers.  To ensure safe immersion of the device during testing, a $2.5~\mu m$-thick layer of SU-8 photoresist was spin-coated onto its surface for waterproofing, and its electrical continuity was verified using a multimeter. It should be noted that conductive arms were added to the ideal electrode pattern to effectively power the device~(see Fig.\ref{Fig2:MEMS}(b)).
The transducer is placed on an acrylic backing layer (cylinder-shaped with the diameter
of 13.5 cm and the thickness of 1.2 cm) and secured in position by a bracket fastened with screws and nuts. The connection wire type between the transducer and the power amplifier is BNC male to wire leads. The wire leads are glued to the electrodes of the transducer by conductive silver paste (Model Z-6011). For waterproof, an insulating glue (Model J-2105) is applied over the connection point.

\subsection{Acoustic Field Measurement}

\begin{figure}
\includegraphics[width=0.75\linewidth]{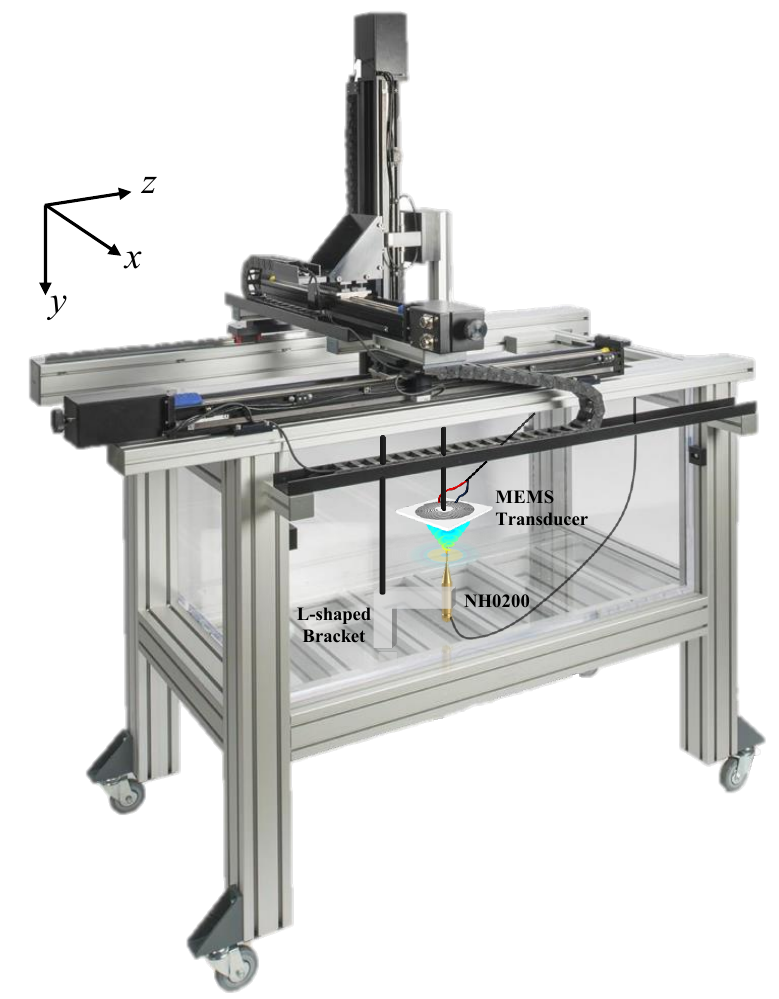}
\caption{\label{Fig3} (color online) Schematic of the experimental setup for acoustic field scanning measurement. Details of the dry-side instrumentation are omitted here for brevity. The coordinate system used in this paper is defined in the upper left corner. The acoustic field scanning follows the procedure of first coarse scanning and then fine scanning. For more details, please see the main text.}
\end{figure}

 As illustrated in Fig.\ref{Fig3}, acoustic field scanning measurements were conducted in a $1 \text{m}\times0.5\text{m}\times0.5\text{m}$ water tank (UMS-4, Precision Acoustics, Dorchester, UK). A custom-developed focused transducer based on MEMS emitted acoustic waves downward into the tank. The acoustic pressure field was measured using a needle hydrophone (NH0200, Precision Acoustics, Dorchester, UK) positioned near the focal region. The hydrophone was mounted on a high-precision three degrees of freedom stepper motor system, allowing controlled scanning of the acoustic field. The received voltage signals were conditioned using a DC coupler (DCPS, Precision Acoustics, Dorchester, UK) and a booster amplifier (Booster Amplifier, Precision Acoustics, Dorchester, UK), and subsequently recorded by an oscilloscope (DSOX3024G, Keysight, Santa Rosa, CA, USA).

 The acoustic field was measured using a two-stage scanning strategy consisting of a coarse scan followed by a fine scan. The coarse scan was first performed over a relatively large spatial region with a step size of approximately $\lambda/3$, where $\lambda$ is the acoustic wavelength in water at the excitation frequency. This preliminary scan was used to verify that the generated acoustic field exhibited the expected structure and to determine an appropriate spatial range for detailed measurements. 
 Subsequently, a fine scan was carried out within a smaller region surrounding the focal area (1~mm $\times$ 1~mm) with a higher spatial resolution. In this stage, the scanning step size was reduced to approximately $\lambda/15$ to accurately resolve the spatial features of the acoustic field. The recorded time domain signals were processed using fast Fourier transform (FFT) to extract the complex acoustic pressure at the driving frequency. The resulting data were then interpolated and smoothed to obtain the spatial distributions of the acoustic field amplitude and phase.
 
\section{\label{sec:results and discussion} Results and discussion}
The acoustic fields generated by the fabricated transducers were measured in a water tank and compared with the numerical results obtained using the angular spectrum method. In the following, we first present the results for the focused vortex device and then for the focused beam device. In both cases, good agreement is observed between experiment and simulation in both the focal plane and the propagation plane.

\subsection{\label{subsec:focused vortex}Experimental and numerical results of the focused vortex}
\begin{figure*}
\includegraphics[width=15.6cm]{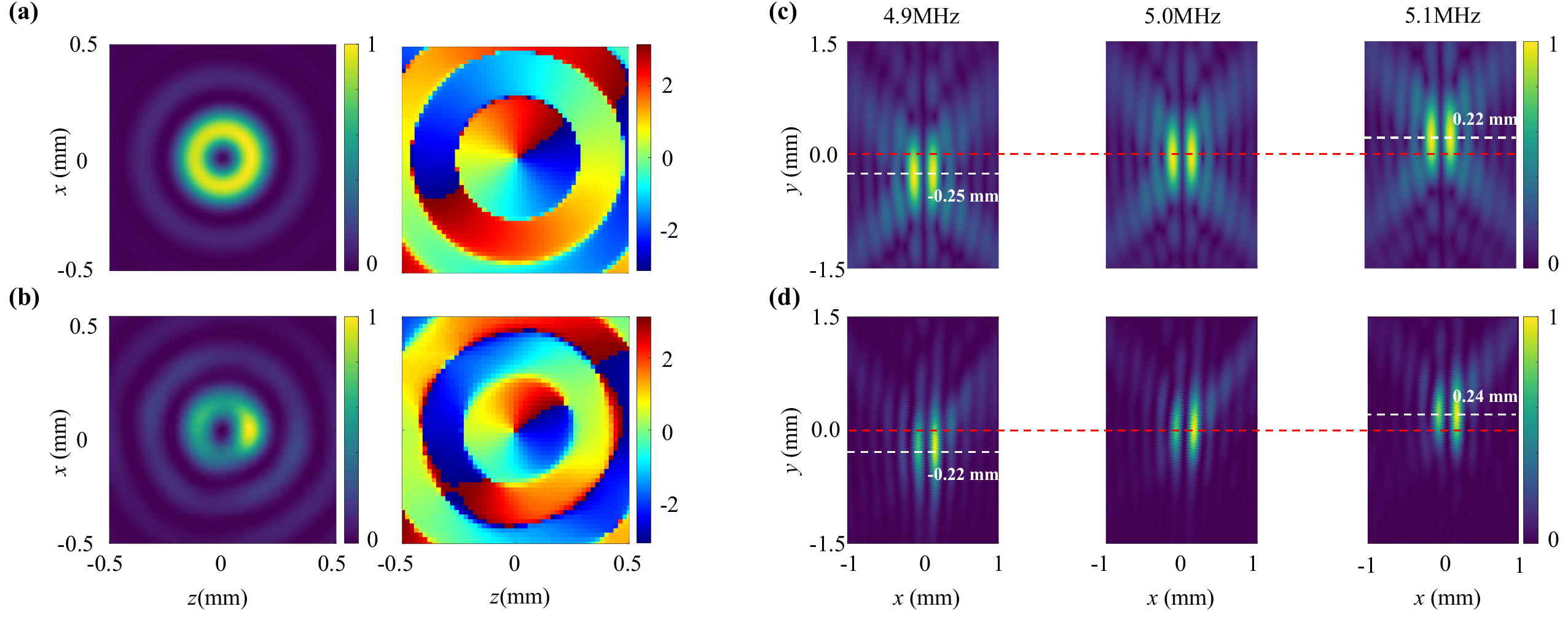}
\caption{\label{Fig4} (color online) Simulated and measured results of the focused vortex. (a) Simulated distributions of acoustic field magnitude and phase in the focal plane ($h_0$= 6 mm) at the designed frequency of 5 MHz. (b) Measured acoustic field magnitude and phase distributions corresponding to the theoretical case in (a). (c) Simulated propagation plane at driving frequencies of 4.9 MHz, 5.0 MHz, and 5.1 MHz. (d) Measured acoustic field results at the corresponding driving frequencies. The red dashed line marks the designed focal length, the white dashed line indicates the actual focal length at the corresponding frequency, and the white numbers indicate the deviation of the actual focal length from the designed focal length. For driving frequencies of 4.9 and 5.1 MHz, the measured focal length changes are -0.22 mm and 0.24 mm, respectively, while the theoretical predictions are -0.25 mm and 0.22 mm.}
\end{figure*}

Fig.\ref{Fig4}(a) and Fig.\ref{Fig4}(b) present the simulated and measured acoustic fields of the focused vortex transducer in the focal plane. The vortex field exhibits a ring-shaped intensity distribution with a low amplitude core at the center, which is the hallmark of a vortex beam. The measured field reproduces this annular structure well, including the ring radius and the overall intensity profile.
The corresponding phase distribution shows a continuous azimuthal variation around the beam center, consistent with the helical phase structure of a focused acoustic vortex. The measured phase agrees closely with the theoretical prediction, confirming that the transducer successfully imposes the intended vortex phase.
The observed $2 \pi$ phase winding indicates a first-order vortex ($m$ = 1).
The propagation plane fields of the vortex beam, shown in Fig.\ref{Fig4}(c) and Fig.\ref{Fig4}(d), reveal a frequency-dependent shift of the focal region. As the driving frequency changes from 4.9 to 5.1 MHz, the location of the high-intensity vortex region moves along the axial direction. The measured results reproduce the simulated field evolution over the entire frequency range, further validating that the fabricated device possesses favorable frequency-tuning focusing characteristics.

Overall, the good agreement between simulation and experiment in both the focal plane and the propagation plane confirms that the proposed transducer design accurately produces the targeted focused vortex. We next show that the same design strategy can be extended to the generation of focused beam.

\subsection{\label{subsec:focused beam}Experimental and numerical results of the focused beam}

\begin{figure*}
\includegraphics[width=15.6cm]{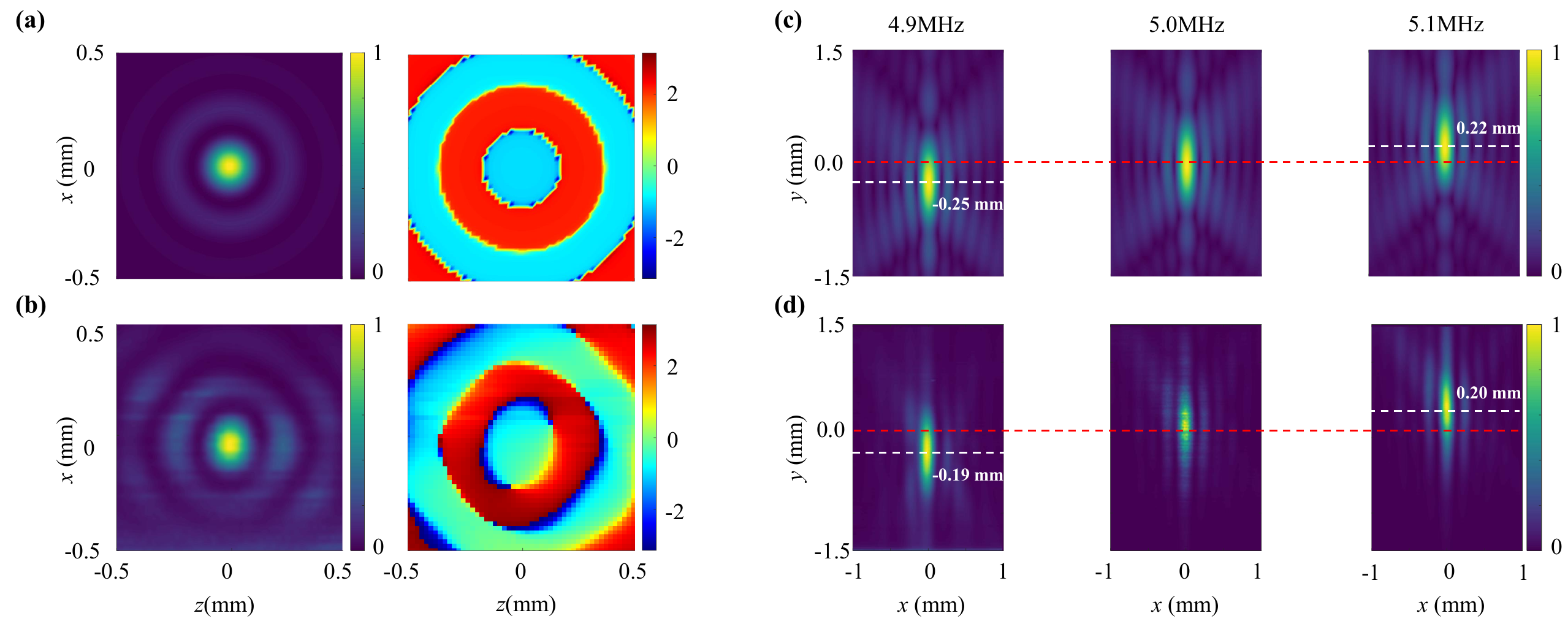}
\caption{\label{Fig5} (color online) Simulated and measured results of the focused beam. (a) Simulated distributions of acoustic field magnitude and phase in the focal plane ($h_0$= 6 mm) at the designed frequency of 5 MHz. (b) Measured acoustic field magnitude and phase distributions corresponding to the theoretical case in (a). (c) Simulated propagation plane at driving frequencies of 4.9 MHz, 5.0 MHz, and 5.1 MHz. (d) Measured acoustic field results at the corresponding driving frequencies. The red dashed line marks the designed focal length, the white dashed line indicates the actual focal length at the corresponding frequency and the white numbers indicate the deviation of the actual focal length from the designed focal length. For driving frequencies of 4.9 and 5.1 MHz, the measured focal length changes are -0.19 mm and 0.20 mm, respectively, while the theoretical predictions are -0.25 mm and 0.22 mm.}
\end{figure*}

Fig.\ref{Fig5}(a) and Fig.\ref{Fig5}(b) compare the simulated and measured acoustic fields of the focused beam transducer in the focal plane. The calculated field exhibits a pronounced central maximum with nearly circular symmetry, which is characteristic of a conventional focused beam. The measured field reproduces the same main-lobe structure and focal spot size, indicating that the fabricated transducer generates the desired focusing pattern with high fidelity. The measured phase distribution also agrees well with the simulated result, confirming that the fabricated focused beam transducer successfully achieves the expected conventional focused wave.

The propagation plane results in Fig.\ref{Fig5}(c) and Fig.\ref{Fig5}(d) further demonstrate the frequency-dependent focusing behavior. As the excitation frequencies are varied from 4.9 to 5.1 MHz, the focal position shifts systematically along the propagation direction as similar as the focused vortex case. The measured field evolution follows the same trend as the angular spectrum simulation.
This observation provides direct experimental evidence for the frequency-tunable focal length predicted by the theoretical model.

\subsection{Comparison and discussion}
In Sec.\ref{subsec:focused vortex} and Sec.\ref{subsec:focused beam}, for both focused vortex and focused beam, the measured acoustic fields show good agreement with theoretical results and the transducer’s frequency‑tuning focusing characteristic was preliminarily validated. The observed discrepancies, such as the non-circular shape of the measured field, are mainly attributed to the conductive arms introduced in the fabricated device. These arms break the structural symmetry of the ideal transducer model, whereas the theoretical calculations assume a perfectly symmetric electrode pattern. Additional minor deviations may also arise from the finite aperture of the hydrophone and alignment errors during scanning. 

\begin{figure}
\includegraphics[width=0.45\textwidth]{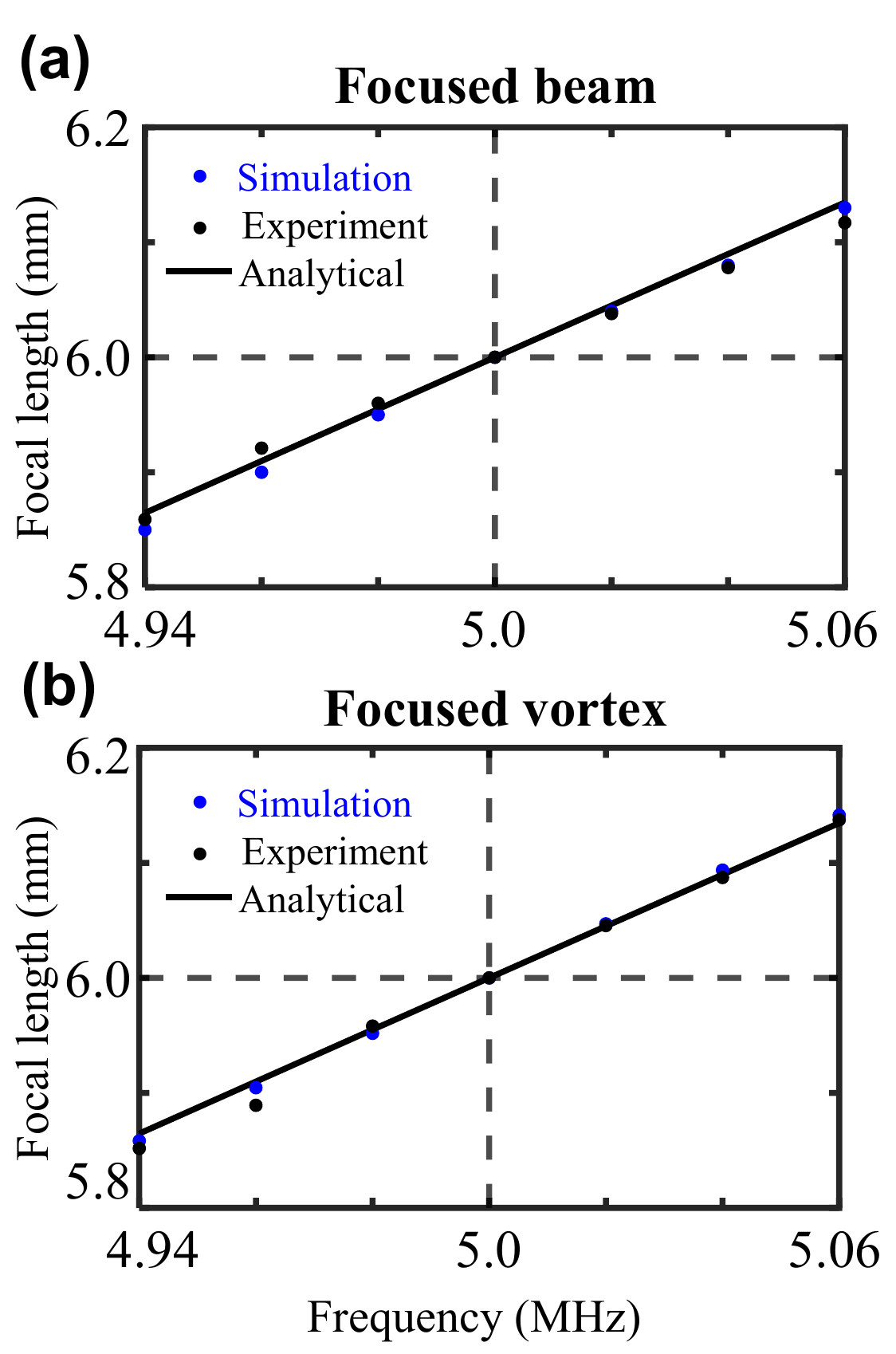}
\caption{\label{Fig6:Theory plot} (color online) Comparison of theoretical, simulated, and experimental focal lengths for different excitation frequencies of the two focused transducers with the design frequency at 5 MHz. (a) Focused beam. (b) Focused vortex.}
\end{figure}

 To further explore the frequency tuning capability of the transducer, multiple acoustic field scanning experiments were performed with varying driving frequencies near the designed frequency, and the focal lengths from theoretical predictions, simulations, and experimental measurements were compared.
Fig.\ref{Fig6:Theory plot} presents the focal length as a function of excitation frequency for the focused beam and the focused vortex. Note that to save scanning time and facilitate measurements, the acoustic field was uniformly scanned at a fixed plane $y$=6 mm with a sampling interval of $\lambda/3$.  The measured data were then interpolated onto a finer grid ($\lambda/15$), and the source-plane field was reconstructed using the backward angular spectrum method. The propagation-plane field was subsequently reconstructed via forward angular spectrum propagation, from which the focal length was extracted from the measured data. 
  In both cases, the focal length increases approximately linearly with frequency over the investigated range around the design frequency. The analytically predicted curves (using Eq.\eqref{Eq:linear equation}) agree closely with the results obtained from the angular spectrum method and with the experimentally measured focal positions. In particular, at the design frequency of 5.0 MHz, both transducers produce a focal length of approximately 6.0 mm, in accordance with the design target.

The close agreement among analytical, simulated, and experimental results confirms the validity of the approximate formula derived for the focal length variation near the design frequency. It also demonstrates that the angular spectrum method accurately captures the finite-aperture propagation of the emitted acoustic field and that the fabricated MEMS transducers realize the intended acoustic field with high fidelity.

Next, we explain why focusing can still persist when the excitation frequency is detuned from the design value $f_0$. Once $f\neq f_0$, the aperture is no longer globally phase matched, so perfect in-phase superposition over the whole source is lost. Nevertheless, the field does not immediately cease to focus. The reason is that each annular region of radius $r$ still contributes constructively around its own local stationary-phase axial position.

To see this, we write the exact phase associated with a source point at radius $r$ and an axial observation point $y$ as
\begin{equation}
    \Phi(r,y)=k \sqrt{r^2 +y^2}-k_0 \sqrt{r^2 + h_0^2} +C,
\end{equation}
where $C$ is a constant and $\alpha=f/f_0=k/k_0$. Using the stationary-phase condition (i.e., identifying the stationary point of the phase with respect to radius~\cite{arfken2011mathematical}),
\begin{equation}
    \frac{\partial \Phi}{\partial r}=0,
\end{equation}
gives
\begin{equation}
    y(r)=\sqrt{\alpha^2 h_0^2 + r^2 (\alpha^2-1)}.
\end{equation}
This expression reveals the physical origin of frequency-tunable focusing. At the design frequency ($\alpha=1$), $y(r)$ is independent of $r$ and reduces to $y=h_0$, meaning that all radii share the same axial stationary-phase position. In this special case, the field is formed by true global in-phase superposition, corresponding to ideal focusing. After frequency detuning ($\alpha\neq1$), however, the stationary-phase position becomes radius-dependent: different radii no longer correspond to exactly the same axial point, but each annular region still retains its own local axial position of strongest constructive contribution. Therefore, the focus does not disappear abruptly. Instead, a finite focal region is preserved as long as the spread of $y(r)$ across the aperture remains small.

Near the design frequency, $\alpha$ is still close to unity, so the variation of $y(r)$ over the aperture is limited. As a result, the local stationary-phase positions of different radii remain clustered in a narrow axial interval, and the field still exhibits a well-defined effective focus. With increasing detuning, the spread of $y(r)$ becomes larger, so the contributions from different radii are distributed over a wider axial range. This leads to progressive defocusing.

\section{\label{sec:conclusion} Conclusions}

In this work, we investigated a planar MEMS interdigital transducer that is optically transparent, facilitating integration with microscopic platforms. The transducer is capable of generating a focused vortex and focused beam, and enables focal tuning through frequency adjustment. The transducer was designed by forming a binary phase hologram, and the corresponding device was fabricated using a MEMS process. The acoustic fields generated by the device were characterized experimentally by water tank scanning in focal plane and propagation plane.

An approximate linear relationship between the focal length and the excitation frequency near the design frequency was derived analytically. Furthermore, a theoretical explanation is provided to elucidate the physical mechanism that enables effective focusing even when the driving frequency is varied within a certain range. The relationship between the focal length and the driving frequency obtained from theory, simulation, and experiment shows excellent consistency, confirming that the fabricated transducer possesses stable frequency-tuning focusing capability and validating the reliability of the proposed theoretical model and simulation framework.

These results demonstrate that frequency tuning provides a simple and effective way to adjust the focal position without mechanical translation or complex phased-array control. In addition, the proposed flat transducer architecture is compatible with low cost fabrication and batch production, which is attractive for compact and integrated ultrasonic systems.

The present study provides a practical framework for the design of frequency-tunable focused acoustic transducers and may be useful for applications requiring controllable acoustic focusing, such as high-resolution imaging, microparticle manipulation, and integrated acoustofluidic devices.


\begin{acknowledgments}
We thank Dr. Michael Baudoin, Dr. Ravinder Chutani, Dr. Roudy Al-Sahely for helpful discussions on the fabrication procedures of the devices.
Z. Gong thanks for the support from the National Natural Science Foundation of China (24Z990200542 and No. 12504522), the Joint Training Fund Project of Hanjiang National Laboratory (No. LP2024004), the XIAOMI Foundation, and the Shanghai Jiao Tong University [2030 Initiative, AI for Engineering Initiative, and the startup funding (WH220401017, WH22040121)].
\end{acknowledgments}

\section*{\label{sec 5}AUTHOR DECLARATIONS}
\textbf{Conflict of Interest} \\
The authors have no conflicts to disclose.

\section*{\label{sec 6}DATA AVAILABILITY}
The data that support the findings of this study are available from the corresponding author upon reasonable request.

\bibliography{main}

\end{document}